\documentstyle[aps,preprint,tighten]{revtex}
\def\be{\begin{equation}}
\def\ee{\end{equation}}
\def\bea{\begin{eqnarray}}
\def\eea{\end{eqnarray}}

\begin{document}
\draft


\title {\bf Ward Identity for Membranes}

\bigskip

\author{A. Ghosh\thanks{e-mail: Amit.Ghosh@cern.ch}}
\address{CERN, Theory Division, CH-1211 Geneva 23, Switzerland}
\author{J. Maharana\thanks{e-mail: maharana@nxth04.cern.ch}}
\address{Max-Planck Institute for Gravitation, Albert-Einstein Institute,
Potsdam, Germany\\ 
and Institute of Physics, Bhubaneswar 751005, India\thanks{permanent address}}
\date{\today}
\maketitle

\begin{abstract}

\noindent Ward identities in the case of scattering of antisymmetric 
three form RR gauge fields off a D2-brane target has been studied in 
type-IIA theory.

\end{abstract}
\vspace{2 cm}

\narrowtext

\newpage


Recent  progress  in  understanding the
nonperturbative aspects of  string theory \cite{asen,w} has provided a
unified description of the dynamics of the five (perturbatively) consistent
 superstring theories. The  
 p-branes and D-branes, which appear naturally as solutions of 
 the string effective action, have a crucial role to play 
in these developments \cite{jp}.     
 It  is believed that there is an underlying fundamental theory and  the five 
string theories  are various phases of this theory
 \cite{jhs,pk1} and the low energy effective action of the fundamental theory, 
named M-theory, can be identified with that of  $D=11$ supergravity.  There is 
mounting evidence that M-theory encompasses and unifies string theories and  
governs string dynamics in diverse dimensions. 
 The bosonic sector of the low energy effective action
of this
theory contains the graviton and the antisymmetric 3-rank tensor and 
consequently, the membrane that  couples to the three index 
antisymmetric tensor field is expected to be the natural extended fundamental 
object \cite{bts,mjd} in eleven dimensions.   
It is natural to explore   
 whether the fundamental 
(super)membrane can provide the degrees of freedoms of M-theory. 
It is not evident that there exists  a consistent 
    quantum mechanical theory of the (super)membrane 
\cite{nicol}.   
This issue is very closely related with the 
large  $N$ behaviour of the $U(N)$ matrix model \cite{gj}. 
The proposal
of the M(atrix) model \cite{matrix} has led to very interesting developments 
\cite{rev2}.
The M(atrix) theory reveals various salient features of the M-theory. In this 
approach, the dynamics of the eleven
dimensional M-theory is described by the many body quantum mechanics
of $N$ $D0$-branes of the type IIA theory in the limit $N \rightarrow \infty$.
The compactified M(atrix) theory gets related to  super
Yang-Mills theories through dualities. Furthermore, it 
provides a theoretical basis to the 
understanding of the microscopic dynamics of M-theory and holds the  promise of
exploring various aspects of string theories nonperturbatively \cite{revm}.

Since the membrane theory provides an intimate connection with M-theory,
there have been attempts 
to study the supermembrane action in curved space with antisymmetric 
tensor field background 
 \cite{dpp}   
 in spite of the above mentioned shortcomings. 
It is interesting to investigate  
 how much the world volume theories know about
the spacetime \cite{pk2}. One of us \cite{jmm} has investigated how the
bosonic membrane theory is encoded with the target space local symmetries such
as general coordinate transformation and gauge transformation; the former
being associated with the graviton and the latter with antisymmetric tensor
field. 
The strategy followed to expose these symmetries was based on  
 the one proposed by Veneziano \cite{g}, in the context of 
string theory, to derive  gravitational Ward identities, and by Veneziano and 
one of the present author \cite{mv,mv1} 
  to derive  Ward identities for various massless excitations of
strings: both compactified and noncompactified \cite{m}. 
Let us recall the technique adopted in ref.[16-18].
A  Hamiltonian phase space framework is adopted to obtain a  
Hamiltonian form of the action. Next,  a set of generating functionals 
associated with the local (target space) symmetries of the theory are
introduced. It  was explicitly demonstrated  that the variation of the action, 
under these canonical transformations,
 can be compensated by appropriate  (gauge) transformation
of the massless backgrounds. The Ward identities are derived with the
argument that the Hamiltonian path integral measure remains invariant under 
canonical transformations, at the classical level at least.

A similar approach was adopted in the case of the membrane  and it was 
  shown  that it is possible
to introduce canonical transformations associated with general coordinate
transformation and gauge transformation in the target space of the M-theory.  
Of course, the  results for the membrane are to be understood as classical one in view of the preceding
remarks regarding the quantum theory of membranes. It is worthwhile to mention
that the the Ward identities  derived for membrane \cite{jmm} are not easy
to verify explicitly, unlike the Ward identities in string theory where
one can obtain explicit expressions for the vertex operators for 
some simple backgrounds and the conformal invariance imposes constraints on the
form of various vertex operators. In the covariant formulation, the vertex operators
are required to be BRST invariant.

The purpose of this note is to derive Ward identities for scattering of the
antisymmetric tensor gauge field from a membrane, specifically D2-brane, that
appear in the type IIA string theory. In fact, the type IIA theory is of
special significance since the strong coupling limit of this theory in 
intimately related to the $D=11$ supergravity \cite{ed}. 
Thus the membrane appearing
in type IIA theory is very closely connected with the membrane of the
eleven dimensional theory. We shall see that our test of Ward identity will be
at the level of scattering amplitudes similar to the works in [16-18], 
although the techniques will be slightly different. 
Let us recall how the Ward identities  are derived in the
string theory [16-18]:

\bea Z[J]=\int[{\rm{dX~dP}}][{\rm {d{\cal G}}}]{\rm exp}~(-iS_H[X,P,{\cal G},
J])\eea
here, $Z[J]$ is the generating functional, $X$ represents the string 
coordinate, $P$ represents the conjugate momentum, ${\cal G}$ are the ghosts
 and $S_H$ is the Hamiltonian action which is
a functional of coordinates, momenta, ghosts and the massless background 
field $J$, 
which corresponds to graviton, antisymmetric tensor or gauge potential
according to the case at hand. As mentioned earlier, one introduces a suitable
generator of the canonical transformation, $\Phi _J [X, P]$. Then it was shown
explicitly that for the above backgrounds, the following relation holds:
\bea \delta _{\Phi _J} S_H = \delta ^J_{Gauge} S_H \eea
In other words, first one computes the variations of $X$ and $P$ under 
$\Phi_J$ to obtain the variation of $S_H$. Then, one checks from the 
$\rm {l.h.s}$ of the above equation that it is the same as implementing 
general coordinate transformation or
Abelian gauge transformation associated with graviton or antisymmetric tensor
field. In the case of compactified string the nonabelian gauge 
fields are also permissible backgrounds and in that case the corresponding
nonabelian gauge transformation is to be considered in the $\rm {r.h.s}$ of 
the equation. Now the argument is that the Hamiltonian phase space measure 
remains invariant under $\Phi _J$ and the variation of $S_H$ under $\Phi_J$ 
can be compensated by appropriate gauge transformation of background $J$ 
leading to the relation
\bea 0=\delta^J_{Gauge}Z[J]=\langle\int d^Dx\left[-i{\delta S_H\over\delta
J(x)}\delta^J_{Gauge}J(x)\right]\rangle _{J_{bg}}\eea
Here $\langle...\rangle$ means that the expression is averaged with
the path integral factor $\int [{\rm dXdPd}{\cal G}]\;{\rm exp}
(-iS_H)$
The interpretation of the above equation is as follows: from the preceding
arguments $Z[J] = Z[J + \delta ^J_{Gauge}J]$ leading to the equation (3). Here 
$J_{bg}$ means that the massless fields $G_{\mu\nu}, B_{\mu\nu}$ or $A_{\mu}$
takes their background values after the functional derivative of the Hamiltonian
action is taken. Notice that
\bea {{\delta S_H}\over {\delta J(x)}} =\int d^2 \sigma \delta (x-X(\sigma))
{{\delta {\cal L}_H} \over {\delta J(X)}} = \int d^2 \sigma \delta (x-X(\sigma))
V_J(X,P) \eea   
where, $V_J(X,P)$ is the corresponding vertex operator for $G_{\mu\nu}, B_{\mu
\nu}$ or $A_{\mu}$ depending on what type of WI one is interested in.
As an example, let us consider a quick derivation of the gravitational WI,
note that  
\bea \delta ^{GCT} G_{\mu\nu} = -G_{\mu\lambda}\xi^{\lambda},_\nu 
- G_{\nu\lambda} \xi^{\lambda},_\mu - G_{\mu\nu},_\lambda \xi^{\lambda} \eea 
Using (3) and (4) we arrive at
\bea \langle \int d^2\sigma \left [V^{\mu\nu}_G (X,P)\{ G_{\mu\lambda}(X)
\xi^{\lambda}(X),_\nu+G_{\nu\lambda} (X) \xi^{\lambda}(X),_\mu + 
G_{\mu\nu}(X),_\lambda \xi^{\lambda}(X)\}\right]\rangle_{G_{bg}} = 0 \eea
here $\xi^\lambda(X)$ is the local parameter associated with infinitesimal
general coordinate transformation. Since it is an arbitrary parameter, if we 
differentiate the above equation with respect to $\xi ^{\lambda}$ and set
 $\xi =0$ the $\rm {r.h.s}$
will be still zero.
Furthermore, we can take functional derivatives of the whole expression 
with respect to a string of G's and set eventually the metric to be flat space
metric (for simplicity) to arrive at 
\bea {\delta^n\over\delta G_{\mu_1\nu_1}(y_1)...\delta G_{\mu_n\nu_n}(y_n)}
\langle\int d^2\sigma V^{\mu\nu}_G \{ G_{\mu\lambda}
\partial_\nu\delta(x-X)&&+G_{\nu\lambda}\partial_\mu\delta(x-X)\nonumber\\
&&+G_{\mu \nu},_\lambda \delta (x-X)\} \rangle _{G=\eta} =0 \eea

Note that the functional derivatives of metric act in three ways: first when
it acts on the path integral factor buried in $\langle ... \rangle $ it brings
down the vertex operator $V^{\mu\nu}_G(X)$, in this case; second, if there is
any G dependence in the vertex operator in the above expression, it removes that
G and introduces a factor of $\delta (y_i -X)$ and thirdly it kills the factor
of G which exists inside the curly bracket. The WI is rather transparent if one
takes the Fourier transform. The derivatives of delta function will give factors
of momenta. Thus $(n+1)$-graviton amplitude gets related to lower point 
amplitudes. We also know that BRST invariance will impose constraints on the
vertex operators.

We shall adopt following prescription to  derive WI for the scattering of three 
index antisymmetric tensor field of type IIA theory
from the D2-brane. Notice that the three-index potential (whose field strength
is four index antisymmetric tensor) comes from the RR sector.
First, we obtain the vertex operator for the
three index antisymmetric tensor field in the type IIA theory. 
The  amplitude for scattering of the gauge
boson from the D2-brane is obtained using the techniques of conformal
field theory. As is well known the vertex operators must be BRST invariant
and the prescription of deriving them in the covariant formulation of 
superstring was given by Friedan, Martinec and Shenker \cite{fms}.
Finally, one can explicitly check that the WI are
satisfied when the scattering amplitude, after separating out the 
`kinematical' factors, is contracted with the momenta
of the incoming or outgoing gauge bosons. 

 Let us recall that the massless excitations of the type II theory arise
from the product of the left and right moving sectors involving NS-NS and
RR oscillators. We can represent this as
\bea \vert \mu \alpha \rangle _R \times \vert \nu \beta \rangle _L \eea
here $\mu, \nu$ refer to the NS-NS sector and $\alpha , \beta $ 
correspond to the RR sector; the former being spacetime indices take values
0,1...9 and the latter are spinor indices.  The familiar bosonic fields
are graviton, dilaton and the antisymmetric tensor coming from the NS-NS
sector. Furthermore, the bosonic fields originating from RR sector
appear as bispinors in the vertex operator
\bea V_{RR}=F_{\alpha \beta} U^{\alpha\beta} \label{vertex}
\eea
the bispinor $F_{\alpha\beta}$ can be expanded in terms
of a complete basis of the ten dimensional gamma matrices (antisymmetric 
products) as follows
\bea F_{\alpha\beta}=\sum_{k=0}^{10}{i^k\over k!}F_{\mu_1\mu_2...\mu_k}
(\gamma^{\mu_1...\mu_k})_{\alpha\beta}\eea
where the k-dimensional tensor $\gamma^{\mu_1...\mu_k}$ is constructed from 
the ten dimensional gamma matrices and it
is antisymmetric with respect to all its indices.
Therefore, the tensors $F_{\mu_1...\mu_k}$ appearing the above equation are
antisymmetric in their indices and one concludes that the massless RR fields 
are antisymmetric Lorentz tensors. As the bispinors have definite chirality
projections, thus all the components of the F's are not independent. As a 
consequence,  type IIA theory has only field strengths corresponding to 
even integers of $k$ and type IIB contains field strengths with odd 
integers of k. The former admits only even branes and the latter only odd 
ones, as is well known.

In the covariant formulation of superstring \cite{fms} the vertex operators
involving RR fields contain the spin field $S^{\alpha}$, the bosonized
ghost $\phi$ and of course the `plane wave' piece $e^{ik.X}$. Their 
combinations have to be such that the vertex operator commutes with the 
BRST charge. Generically, we can write
\bea U_{\alpha}^{\beta} = V_{(-{1\over 2}) \alpha}(z){\bar V}^{\beta}_
{(-{1\over 2})}
(\bar z)  \eea
with
\bea V_{(-{1\over 2})\alpha}(z) = e^{-{1\over 2}\phi (z)}S_{\alpha}(z)
e^{ikX(z)} \eea
Note that the bar on the argument of the vertex operator refers to complex
conjugation here and everywhere.
Moreover, the $\gamma$ matrices are  $32 \times 32$ dimensional  and
have the representation

\bea \gamma ^{\mu} = \left( \begin{array}{cc} 0 & \gamma ^{\mu \alpha \beta} \\
\gamma ^{\mu}_{\alpha \beta} & 0 \end{array} \right) \eea
satisfying the anticommutation relation $\{\gamma ^{\mu} , \gamma ^{\nu} \} = -2
 \eta ^{\mu \nu}$ with $\eta ^{\mu \nu} = {\rm diag} ~ (-1,1...1)$; furthermore, 
$\gamma^{11} = \gamma ^0...\gamma ^9 = {\rm diag}\left(1,-1\right)$. 
 The standard method for the 
construction of  $S^{\alpha}$ is \cite{fms} to bosonize the worldsheet fermions,
and introduce a cocycle operator.  In the case of D-brane some of the
coordinates (and therefore also worldsheet fermions) satisfy Dirichlet or
Neumann boundary condition. Thus $S(z)$ and ${\bar S}(\bar z)$ get related
depending on the Dp-brane one is considering. In our case, IIA,
\bea {\bar S}^{\alpha}(\bar z) = M^{\alpha \beta}S_{\beta}(z) \eea
with $M = \gamma ^0 ....\gamma ^p $ for Dp-brane which are even.

Let us consider scattering of massless R-R 3-form states in Type-IIA off
a D2-brane target. In components form the 4-form field-strength is given 
by 
$$F_{\mu\nu\rho\lambda}=\partial_\mu C_{\nu\rho\lambda}-
\partial_\nu C_{\rho\lambda\mu}+\partial_\rho C_{\lambda\mu\nu}-
\partial_\lambda C_{\mu\nu\rho}$$
where $C_{\mu\rho\lambda}$ is the 3-form potential. 
Now, in the absence of the antisymmetric tensor field $B_{\mu\nu}$ the 
Chern-Simons like term $(F\wedge F\wedge B)$,
where $B$ is the NS-NS 2-form potential, is not present in the low energy
effective action; therefore, the equation of motion for the RR field 
strength is given by
\bea\partial^\mu F_{\mu\nu\rho\lambda}=0\eea
Now we choose a plane wave ansatz for the gauge potential 
\bea C_{\mu\nu\lambda}&=&
\epsilon_{abc}\varepsilon^a_\mu\varepsilon^b_\nu\varepsilon^c_\lambda
e^{ik\cdot X},\qquad a,b,c=1,2,3;\quad\epsilon_{123}=+1\nonumber\\
F_{\mu\nu\rho\lambda}&=&i\epsilon_{abc}[
k_\mu\varepsilon^a_\nu\varepsilon^b_\rho\varepsilon^c_\lambda-
k_\nu\varepsilon^a_\rho\varepsilon^b_\lambda\varepsilon^c_\mu+
k_\rho\varepsilon^a_\lambda\varepsilon^b_\mu\varepsilon^c_\nu-
k_\lambda\varepsilon^a_\mu\varepsilon^b_\nu\varepsilon^c_\rho]e^{ik
\cdot X}\eea
where, $k$ is the momentum, satisfying the on-shell condition $(k^2=0)$;
 and $\epsilon^a_\mu$ are
polarization vectors subject to constraints $\epsilon^a\cdot k=0$ as a 
consequence of eq.(15).
 The  amplitude for scattering of the three index gauge field off the D2-brane
involves computation of the correlation functions involving two of the
$V_{RR}$ operators:
\bea
A=\int{dz_1d\bar z_1dz_2d\bar z_2\over(Vol)_{conformal}}\langle V_1(
z_1,\bar z_1)V_2(z_2,\bar z_2)\rangle\;,\label{ampli}
\eea
where $(Vol)_{conformal}$ represents the conformal group volume factor that
has to be factored out and $V_1,V_2$ correspond to the vertex operators of 
asymptotically incoming and outgoing massless R-R states. The precise form
of these vertex operators for this case are given below 
\bea
V={1\over(4!)^2}F_{\mu_1\mu_2\mu_3\mu_4}U^{\alpha \beta}(\gamma^{[\mu_1}
\gamma^{\mu_2}\gamma^{\mu_3}\gamma^{\mu_4]})_{\alpha \beta}\;.
\eea
The computation of the correlation function in (\ref{ampli}) involve the
correlators of four spin-fields
$V_{-{1\over 2}\alpha}(z)$, $V_{-{1\over 2}\beta}(\bar z)$, $V_{-{1\over 2}
\gamma}(w)$ and $V_{-{1\over 2}\delta}(\bar w)$, defined through (12),
 with the indices $\alpha, \beta
,\gamma$ and $\delta$ are appropriately contracted with the products
of $F_{\mu\nu\rho\lambda}$ and the gamma matrices (see eq.10). This is
known and has been calculated in \cite{KLL}. The amplitude can be
computed using the techniques of \cite{fms} as was considered by \cite{hk}.
When the gauge potential is taken to be of plane wave form, then the amplitude
is given by  the following expression  
\bea
A={\Gamma(s)\Gamma(t)\over\Gamma(s+t+1)}\left[(s+t)P_1+sP_2\right]
\eea
where, $s=2k^2_{1\|}=2k^2_{2\|}$ and $t=k_1\cdot k_2$, $k_1$ and $k_2$ are
incoming and outgoing momenta respectively of the massless plane wave
states and $k_{i\|}$ are the components of $k_i$ parallel to D2-brane. 
Furthermore, $P_1$ and $P_2$ appearing in (17) are given by the traces of gamma 
matrices.
\bea
P_1&=&{1\over 4}{1\over 4!}{1\over 4!}\;\bigg[
\epsilon_{abc}\big[
k_{1\mu}\varepsilon^a_\nu\varepsilon^b_\rho\varepsilon^c_\lambda-
k_{1\nu}\varepsilon^a_\rho\varepsilon^b_\lambda\varepsilon^c_\mu+
k_{1\rho}\varepsilon^a_\lambda\varepsilon^b_\mu\varepsilon^c_\nu-
k_{1\lambda}\varepsilon^a_\mu\varepsilon^b_\nu\varepsilon^c_\rho\big]
A^{\mu\nu\rho\lambda\sigma}A^{\mu'\nu'\rho'\lambda'\sigma'}\eta_{\sigma
\sigma'}\nonumber\\
&&\epsilon_{a'b'c'}\big[
k_{2\mu'}\varepsilon^{a'}_{\nu'}\varepsilon^{b'}_{\rho'}\varepsilon^{c'}_{\lambda'}-
k_{2\nu'}\varepsilon^{a'}_{\rho'}\varepsilon^{b'}_{\lambda'}\varepsilon^{c'}_{\mu'}+
k_{2\rho'}\varepsilon^{a'}_{\lambda'}\varepsilon^{b'}_{\mu'}\varepsilon^{c'}_{\nu'}-
k_{2\lambda'}\varepsilon^{a'}_{\mu'}\varepsilon^{b'}_{\nu'}\varepsilon^{c'}_{\rho'}\big]e^{i
(k_1+k_2)\cdot X}\bigg]\eea
where, $A^{\mu\nu\rho\lambda\sigma}={\rm Tr}\;[\gamma^\mu\gamma^\nu\gamma^\rho
\gamma^\lambda\gamma^0\gamma^1\gamma^2\gamma^\sigma]$ and
\bea
P_2&=&{1\over 2}{1\over 4!}{1\over 4!}\;\bigg[
\epsilon_{abc}\big[
k_{1\mu}\varepsilon^a_\nu\varepsilon^b_\rho\varepsilon^c_\lambda-
k_{1\nu}\varepsilon^a_\rho\varepsilon^b_\lambda\varepsilon^c_\mu+
k_{1\rho}\varepsilon^a_\lambda\varepsilon^b_\mu\varepsilon^c_\nu-
k_{1\lambda}\varepsilon^a_\mu\varepsilon^b_\nu\varepsilon^c_\rho\big]
C^{\mu\nu\rho\lambda\sigma\mu'\nu'\rho'\lambda'\sigma'}\eta_{\sigma\sigma'}\nonumber\\
&&\epsilon_{a'b'c'}\big[
k_{2\mu'}\varepsilon^{a'}_{\nu'}\varepsilon^{b'}_{\rho'}\varepsilon^{c'}_{\lambda'}-
k_{2\nu'}\varepsilon^{a'}_{\rho'}\varepsilon^{b'}_{\lambda'}\varepsilon^{c'}_{\mu'}+
k_{2\rho'}\varepsilon^{a'}_{\lambda'}\varepsilon^{b'}_{\mu'}\varepsilon^{c'}_{\nu'}-
k_{2\lambda'}\varepsilon^{a'}_{\mu'}\varepsilon^{b'}_{\nu'}\varepsilon^{c'}_{\rho'}\big]e^{i
(k_1+k_2)\cdot X}\bigg]
\eea
where, $C^{\mu\nu\rho\lambda\sigma\mu'\nu'\rho'\lambda'\sigma'}=
{\rm Tr}\;[\gamma^\mu\gamma^\nu\gamma^\rho\gamma^\lambda\gamma^0\gamma^1
\gamma^2\gamma^\sigma\gamma^{\mu'}\gamma^{\nu'}\gamma^{\rho'}\gamma^{\lambda'}
\gamma^0\gamma^1\gamma^2\gamma^{\sigma'}(1+\gamma^{11})]$. 
Now let us look at the expressions for $P_1$ and $P_2$. Each one can be 
written as a product of a tensor involving only trace of the gamma matrices 
times another piece which contains polarization tensors and momenta. Let us
denote them  as $ T^{(1)}_{\mu\nu...\mu'\nu'}$ and 
$T^{(2)}_{\mu\nu...\mu'\nu'}$ for $P_1$ and $P_2$
respectively. It is easy to check that when the tensors $T^{(1)}$ and $T^{(2)}$
are contracted with $k_{1 \mu}$ or $k_{2\mu '}$, the corresponding momenta of 
incoming  and outgoing gauge bosons, then both $P_1$ and $P_2$
vanish separately and therefore the scattering amplitude, $A$, given by 
eq.(17), vanishes. This is
the  gauge invariance of the scattering amplitude reflected through the Ward 
identity.
                               
We recall that when the BRST invariance condition is imposed in the first
quantized version of string theory on its massless backgrounds one obtains
the equations of motion for those backgrounds. 
For example, this requirement, in the NS-NS massless
sector  of the closed string, imposes constraints on the polarization
tensors of graviton and antisymmetric tensor field in addition to the 
mass-shell condition. Similarly, if we consider the scattering of massless RR
fields the BRST invariance restricts the form of the vertex operators as has 
been investigated by Polyakov \cite{pol}. For Abelian $U(1)$ gauge field
he derived the Maxwell equation by imposing BRST invariance on the
corresponding RR vertex operator for vector bosons in type II theory. 
For the case  at hand, the BRST invariance gives rise to constraints on
polarization tensor and mass-shell condition; 
in other words $k^\mu F_{\mu\nu\rho\lambda}=0$ is a sufficient
condition for the BRST-invariance of the vertex operator (9). Furthermore, it
has been shown \cite{pol} that one can calculate three point function,
using the conformal field theory techniques, involving the dilaton and
RR gauge fields. This interaction does not show up in the string effective
action expressed in the string frame metric. However, if one first goes over
to the Einstein frame by a conformal transformation (involving the dilaton)
and the redefines the RR gauge fields, the interaction terms involving the 
dilaton and RR gauge fields can be exhibited and the correspondence with
the three point function mentioned above can be established. In the light
of our investigation we can conclude that the derivation of the Ward
identities for the scattering amplitude of D2-brane and 3-form potential
is a consistency check of the
current conservation as is also reflected in the Ward identities associated 
with conventional gauge theories.

We note that in the conformal field theoretic calculation of the scattering
process involving D2-brane and the three-index antisymmetric gauge fields,
we do not see the effects of the $(F\wedge F\wedge B)$-like term at this
order. The presence of this CS-like term, at the tree level calculations,
can be seen if one looks at the $(B-F-F)$-vertex\footnote{We thank to S.
Mukhi for discussions related to this issue}\,. 

We mention in passing that one might envisage our results from the
perspectives of M theory. It is
well known that if we start with the M-theory membrane and compactify 
one of the transverse directions on a circle we obtain the type-IIA 
D2-brane with equal tension. Furthermore, the antisymmetric
3-form in d=11 supergravity gives rise to the RR 3-form under Kaluza-Klein
reduction. Therefore, the symmetries uncovered by the Ward
identities in the scattering of RR 3-form from D2-brane in IIA theory 
should also be obeyed in the d=11
theory when one considers 3-form membrane scattering amplitude. We
interpret it, at the  present level of our understanding, that this is an
indirect evidence for the abelian gauge invariance in quantized M-theory. 
Another check would be to calculate directly the scattering amplitude
of the 3-form field in eleven dimensional supergravity limit off the M-theory 
membrane as the target. Similar calculations have been done in ten
dimensional supergravity using extreme black p-branes as target and
the amplitude has been compared with the string calculation when the
probe energy is small or the impact parameter is large compared to the
string scale. They seem to agree perfectly in this limit \cite{hk}.
At this stage it is not possible to compute the scattering of three form
gauge field from the membrane in the 11-dimensional theory in a reliable
manner, as compared to the scattering of 3-form gauge field from D2-brane
in the case of type IIA theory using the conformal field theory techniques.
It will be interesting to see whether such perfect matchings come out of
a M(atrix) theory computation. We may mention that the type-IIA
D2-brane being a dynamical object asks for a consistent effective 
quantum description of the M-theory membrane by which one can hope to 
provide direct checks of these symmetries in d=11, most possibly using
similar techniques discussed at the beginning of this paper and in \cite{jmm}.

In view of these results, it will be possible to derive Ward identities for
the entire massless supergravity multiplets of the ten dimensional type II
theories and supergravity and super Yang-Mills theories obtained from
other string theories, 
although a partial results were derived by Veneziano and JM a few    
years ago \cite{sjmgv,more}. We hope to report our results in future.\\ 

\noindent{\bf Acknowledgements}:
We would like to thank Gabriele Veneziano for valuable suggestions during
the course of this work. One of us (JM) would like to acknowledge useful
discussions with H. Nicolai and would like to thank Max-Planck Institute
for Gravitation for warm hospitality.

\end{document}